# Ride-hailing Impacts on Transit Ridership: Chicago Case Study

**Helena Breuer**
Graduate Research Assistant, Center for Sustainable Mobility, VTTI
3500 Transportation Research Plaza
Virginia Tech, Blacksburg, VA 24061
Email: lenab19@vt.edu

**Jianhe Du**
Senior Research Associate, Center for Sustainable Mobility, VTTI
3500 Transportation Research Plaza
Virginia Tech, Blacksburg, VA 24061
Email: jdu@vtti.vt.edu

**Hesham Rakha, Ph.D., P.Eng. (Corresponding author)**
Director, Center for Sustainable Mobility, VTTI
Samuel Reynolds Pritchard Professor of Engineering, Civil and Environmental Engineering
3500 Transportation Research Plaza
Virginia Tech, Blacksburg, VA 24061
Phone: (540) 231-1504, fax: (540) 231-1555, email: hrakha@vt.edu

Submitted for Presentation at the 2021 TRB Annual Meeting.

## ABSTRACT

Existing literature on the relationship between ride-hailing (RH) and transit services is limited to empirical studies that lack real-time spatial contexts. To fill this gap, we took a novel real-time geospatial analysis approach. With source data on ride-hailing trips in Chicago, Illinois, we computed real-time transit-equivalent trips for all 7,949,902 ride-hailing trips in June 2019; the sheer size of our sample is incomparable to the samples studied in existing literature. An existing Multinomial Nested Logit Model was used to determine the probability of a ride-hailer selecting a transit alternative to serve the specific O-D pair, P(Transit|CTA)[1]. We find that 31% of ride-hailing trips are replaceable, whereas 61% of trips are not replaceable. The remaining 8% lie within a buffer zone. We measured the robustness of this probability using a parametric sensitivity analysis and performed a two-tailed t-test. Our results indicate that of the four sensitivity parameters, the probability was most sensitive to the total travel time of a transit trip. The main contribution of our research is our thorough approach and fine-tuned series of real-time spatiotemporal analyses that investigate the replaceability of ride-hailing trips for public transit. The results and discussion intend to provide perspective derived from real trips and we anticipate that this paper will demonstrate the research benefits associated with the recording and release of ride-hailing data.

---

[1] This value defines the replaceability of the trip, where a value ranging from 0 to 0.45 is considered not-replaceable (NR), and a value ranging from 0.55 to 1.0 is considered replaceable (R).

# TERMINOLOGY

The following is an extensive list of terminology relevant to this paper, and their corresponding contextual definitions.

***First- and last-mile (FLM)***: this refers to the first and last leg of the transit trip that connect the individual from their origin to the first transit stop, and/or from the last transit stop to their destination.

***In-vehicle travel time (IVTT)***: this is the portion of the total travel time, and accounts for all time spent traveling inside the transit vehicle(s). In our analysis, we may refer to this as the "transit time".

***Not-replaced (NR) trip/group***: this is the group containing all transit-equivalent trips with a $P(Transit|CTA) \leq 0.45$. A transit-equivalent trip that has a probability in this range [0 ∪ 0.45] is deemed to be inviable to the individual, and ultimately, does not compete with the RH trip service.

***Out-of-vehicle travel time (OVTT)***: this is a portion of the total travel time outside of the vehicle(s), i.e. accessing, egressing, wait time, transfer walk time.

***Pooled Trip, Ride-hailing***: these are ride-hailing trips that combine two or more trips, such that passengers 'share' the ride. In some scenarios, all passengers meet at a specified location and are dropped off at a shared location. Whereas in other scenarios, passengers are picked up at their desired location and then dropped off in the most efficient order.

***Replaced (R) group***: this is the group containing all transit-equivalent trips with a $P(Transit|CTA) \geq 0.55$. A transit-equivalent trip that has a probability in the range [0.55 ∪ 1.0] is considered a viable mode of service for the specific O-D pair.

***Ride-hailing (RH)***: this refers to the act of servicing a trip via a transportation network company (TNC). Users must have an account with the respective TNC, and have the app downloaded onto their smartphone. These trips are ordered using the TNC's app and require the user to input their destination, whereas the origin is automatically determined using the smartphone's internal GIS software. TNC trip fare pricing is dynamic and dependent on the surrounding demand. Ride-hailing trips can be pooled or single passenger, refer to their definitions.

***Route***: refers to the output from ArcGIS' Route Analysis: the transit-equivalent route for a given RH trip.

***Sensitivity Condition:*** these are the percent-changes in the sensitivity variable used for the analysis. For each variable, there existed 20 sensitivity conditions, ranging from -50% to +50% in increments of 5%, where the 0% condition is the observed values and results.

***Single-Passenger Trip, Ride-hailing***: these are RH trips where a single person ordered the trip, and there is one origin and destination. In some cases, these trips can have more than one passenger.

***TNC***: Transportation Network Companies; these are the private businesses that offer ride-hailing services. Examples include Uber and Lyft.

***Total travel time, transit (TTT)***: this is the time elapsed between the departure time at the origin and the arrival time at the destination for the transit-equivalent trip. This value accounts for OVTT and IVTT if applicable.

***Transit-equivalent trip/CTA-equivalent***: this refers to the routing solution from ArcGIS' Route Analysis. It is important to consider that the output trip does not necessarily use transit. Under certain conditions, the program determines that it is most efficiently serviced by walking, thus a "transit-equivalent" trip does not imply the use of transit.

**Walk time (WT):** this is the sum of time allocated to walking and is a portion of the OVTT. This value is output by the ArcGIS Route Analysis, and assumes a walking speed of 5 km/hr.

## INTRODUCTION

The first ride-hailing[2] (RH) service came to US markets in 2008 when Travis Kalanick and Garret Camp established their company, *Uber* [1]. Four years later, an existing carpooling company, *Zimride*, launched a competing ride-hailing service in San Francisco later renaming itself *Lyft*, to exclusively operate as a ride-hailing service [2]. Over the next two years, competition heightened as Uber expanded to 60 cities across six continents, and Lyft announced its plan to expand to 24 more cities, totaling coverage of over 60 cities [2, 3]. As of January 2019, nearly a decade later, 36% of US adults have used or currently use ride-hailing services [4].

Ride-hailing is best defined as an on-demand, app-based, and real-time service that provides customers with door-to-door transportation for a single trip [5]. Through the company's smartphone app, a customer enters a specific pick-up and drop-off location (O-D pair). On the backend, the RH company's algorithm calculates an appropriate route and trip fare. It then selects the optimal driver to service the trip and notifies the customer of the estimated pick-up time and vehicle/driver details.

Coincidentally, when the ride-hailing market began rapidly gaining traction through geographic expansion and increased acceptance in 2014, average public transit ridership in the United States began its decline. In the early twenty-first century, transit ridership in the United States experienced two periods of growth followed by decline (Figure 1).

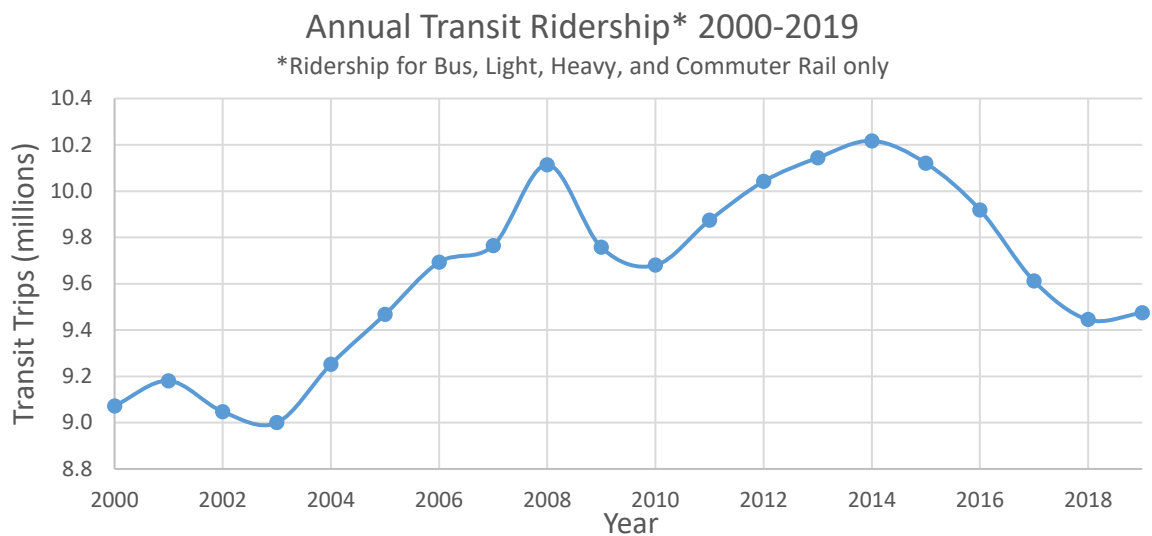

**Figure 1: Annual Public Transit Ridership in the United States from 2000-2019. Source Data: APTA Ridership by Mode and Quarter 1990-Present [6]. Annual ridership counts are the sum of bus, light rail, commuter rail, and heavy rail trips**

On average, from 2003 to 2008, public transit ridership in the United States increased by 2.58% each year. Following the 2008 economic recession, ridership levels took a downturn until 2010 when ridership began increasing again until 2014. Although, unlike the first period of growth, this growth rate was decreasing in magnitude each year until it plateaued in 2014, as illustrated in Figure 2. At this point, transit ridership began rapidly decreasing by losing more riders per year until 2019. While these statistics measure the national trend in mode-choice behavior, the trends within metropolitan transit agencies vary by year and mode. Nonetheless, continuous decline in ridership is significant and likely indicative of a disturbance to the market.

---

[2] Within the existing literature, *ride-hailing* is more commonly known as "ridesharing" but because it entails 'hailing' a ride which is not necessarily shared, *ride-hailing* is most appropriate.

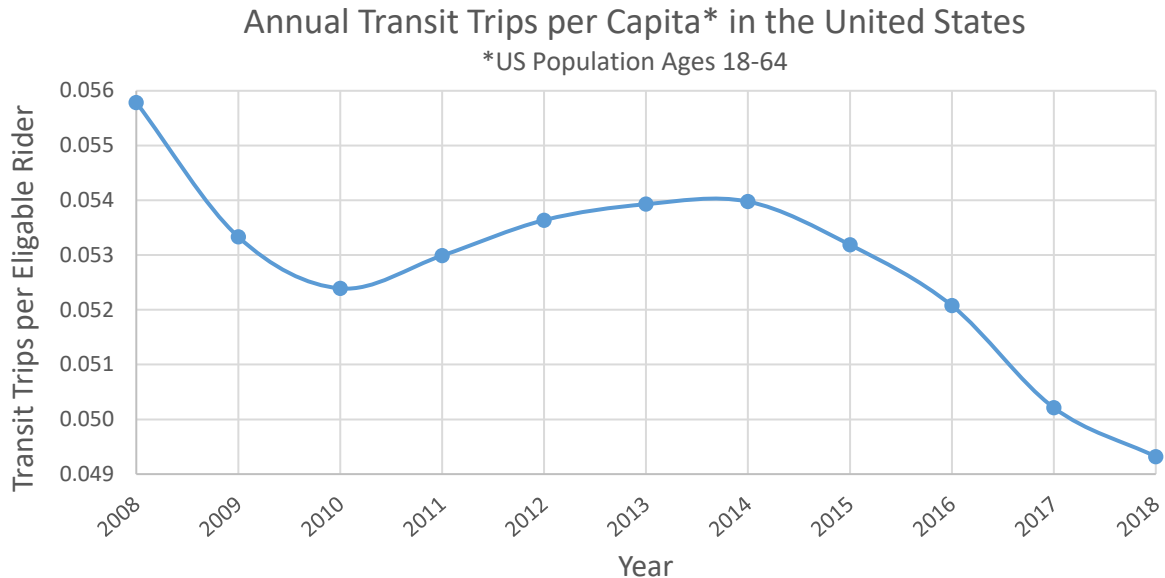

**Figure 2: Annual Transit Trips per Eligible Rider in the United States. Source Data: APTA Ridership by Mode and Quarter 1990-Present, and Population by Age from KFF**

From 2008 to 2018, the population of eligible riders[3] increased by 5.63%. Yet, the annual transit trip per eligible rider decreased by 11.6%, as shown in Figure 2. Moreover, the transit ridership per eligible rider decreased by 8.6% from 2014 to 2018. Despite the steady growth of the US population, transit ridership does not reflect that.

Historically, declines in transit ridership can be a result of macroeconomic, geographic, and demographic changes in a region. The first period of ridership decline in the 21st century started in 2008 and was evidently a ramification of the economic recession. Yet, the cause(s) of the most recent decline is not as discernable. Further, this period exhibited a larger decline in magnitude and has spanned 5 years, as opposed to 2 years. So, what could have possibly caused a more crippling effect on transit ridership than the economic recession? Was the second decline in transit ridership a result of an alternative mode, *ride-hailing*?

This paper will approach answering this question and will fulfill the literature gap by identifying the replicability of transit by RH services through non-empirical based methods. In our study we use ride-hailing trip source data from the City of Chicago containing over 8,000,000 trips, hence the contributions of our research are expanded by the use of a relatively large dataset. We use spatial analyses and methodologies to deliver a real-time transit-equivalent route. We then input trip characteristics into our selected utility model, from which we will calculate the corresponding probability that an original ride-hailer would select transit over RH. We will perform two analyses, the first exploring the replaceability of all RH trips as a whole, and second, a parametric sensitivity analysis of the P(Transit|CTA) with respect to four parameters.

---

[3] Eligible riders: total US populations between age 18 and 64 7.  *Population by Age.* 2018, Kaiser Family Foundation: Online.

## LITERATURE REVIEW

It is important to review the *Terminology* section prior to reading this section, as much existing literature includes new terminology.

The current body of research on ride-hailing is limited by its novelty and the lack of publicly available ride-hailing trip data. Given that ride-hailing services were first introduced to the market in 2010, external research on its utility and impact is relatively untouched. Moreover, ride-hailing services are privately, consequently, trip-specific data is exclusively withheld and unavailable for public research use. While there is no existing literature that definitively states how ride-hailing services impact public transit ridership, many stipulate a correlation between the two, and if ride-hailing is a contributor, it is likely not acting alone.

This absence of trip data has led researchers to obtain empirical data through stated preference (SP) and revealed preference (RP) surveys [8-11]. Some studies executed intercept surveys at points of interest [12], and one executing in-person interviews [13].Yet to our knowledge, there exists no research on the relationship between ride-hailing and public transit that is uses source-data. Consequently, these empirical methods confine the spatial and temporal ranges, limiting the application and testing the integrity of the findings. Ultimately, this has led to conflicting arguments that have yet to be resolved. In the following literature review, we will identify reoccurring themes and findings regarding the impact of ride-hailing services on public transit ridership. We will also highlight the methods used to obtain data. Lastly, we will determine gaps in the literature and how we will address these gaps in our study.

It is important to note that the relationship between ride-hailing and public transit encompasses one-to-many relationships. Bus and rail (light and heavy) both fall under 'public transit', although each mode services trips of differing purposes, rider demographics, and LOS metrics. Accordingly, most literature analyses each modality separately.

In general, the current literature seeks to explore the impact of ride-hailing on VMT and vehicle emissions, its relative safety, and its effect on mode selection. Yet, the latter of the three concerns is the least explored. Contreras and Paz presented three questions, one of which illustrates this concern, "have RHC's [ride-hailing companies] had a negative or positive effect on transit ridership and/or revenue?" [14]. Answering this question requires empirical and source-data based research.

As stated previously, the lack of source-data based research has led to conflicting arguments. Considering that "public transit" encompasses many transit modes, positions tend to be unique per mode (bus, rail). The first position argues that the perceived gains of ride-hailing services attract riders and thereby, substitutes transit. This is based on the significant difference between the gains, and marginal difference between the cost between public transit and ride-hailing. Thus, the cost differential is perceived to be worth the gain of ride-hailing, and thereby replacing public transit. Accordingly, these critics pose that ride-hailing services contribute to the recent decline in public transit ridership. Whereas the second and opposing position argues that ride-hailing complements and reinforces the use of public transit by servicing the first- and/or last-mile (FLM) arrangement, and therefore induces revenue.

Most studies have explored mode choice behavior towards ride-hailing through observation-based research methods, such as SP, RP, and intercept surveys [9, 10, 15]. According to Clewlow and Mishra, ride-hailing services replaced 6% of bus trips and 3% of light rail trips. Whereas ride-hailing acted complementary to commuter rail services, increasing ridership by 3%. Similarly, Graehler et al. found that the entry of a TNC decrease heavy rail and bus ridership by 1.3% and 1.7%, respectively [16].

Rayle et al. determined the primary reasons why individuals chose ride-hailing over the alternative of interest. In brief, users chose ride-hailing over the bus because it was faster and over rail because it was faster, easier to pay, and had less wait time [12].

Henao and Marshall worked as Uber drivers to obtain observational data in real time. Verbal and recorded interviews were used during ride-hailing trips. Of the 311 passengers interviewed, only 5.5% of riders were using the ride-hailing service to get to or from a transit station [13]. This implies that 94.5% of ride-hailing trips do not service the FLM arrangement. However, the small sample size challenges the range of application and the question with a binary response option minimizes bias.

Nelson and Sadowsky used a difference in differences (DID) modeling by comparing transit ridership and operational metrics red before and after the entry of ride-hailing service(s). Their findings concluded that transit ridership increased following the entry of the first ride-hailing company, then decreased once the second company entered the regional market. The presence of the second company led to competition and increased affordability, allowing it to appeal to more people [17].

In 2016, APTA investigated the relationship between emerging modalities and public transit. The research areas included seven major US cities: Austin, Boston, Chicago, Los Angeles, San Francisco, Seattle, and Washington DC. Researchers executed in-depth interviews with transportation officials and surveying of network users. The most relevant finding is shared modes, i.e. ride-hailing services, are used most frequently for social trips during hours which public transit is not in operation or has reduced services. Hence, when transit operations are reduced, ride-hailing services supplement its decreased availability. Results from the survey show that 54% of respondents had used "ride-sourcing" (ride-hailing) to serve a recreational or social trip within the previous 3 months. Further, only 21% of respondents claimed to have used these services for commuting within the previous 3 months. However, this survey does not look at the trend in demand by trip type over a period of time. The percentage of respondents claiming to have used ride-sourcing for a specific purpose does not encapsulate the frequency of demand by type. For example, 21 out of 100 respondents could use ride-hailing services for commuting on a daily basis, whereas 74 out of 100 respondents only used ride-hailing once a week for recreational/social trips. The cumulative demand by trip purpose cannot be represented through a one-time survey [18].

While these methods are useful and highly qualitative, they assume an ideal condition that respondents are not biases. Hence, the results are vulnerable to many biases. The first, *hypothetical bias*, is the propensity of humans to view survey questions hypothetically to an extent that skews responses validity. Second, *strategic bias* is the tendency for a respondent to evaluate their hypothetical behavior such that it favors the response with greater perceived value. Lastly, *framing bias* is how the phrasing and wordage of a question influences its interpretation.

The overwhelming use of surveys and interviews serves as an opportunity to deploy a more quantitative study that focuses on individual trips and their corresponding LOS attributes. Until we can collectively concur upon the effect of ride-hailing, designers, planners, and politicians cannot make sound decisions. We hope to contribute to the field by pioneering new methods and approaches to analyze the impact of RH on public transit ridershiop. Our publication of source data-based research will not only result in greater clarity and insight but will illuminate gray areas with more intensity. From this, empirical studies should be refined to focus on investigating these ambiguous regions and identifying their sources.

## METHODS

### Geography and Demographics of Chicago
Per the U.S. Census, the population of Chicago was estimated to have been 2,705,994 persons in July 2018 across 801 census tracts. From 2014-2018, there was an average of 1,056,118 households with a median income of $57,238 [19]. As of 2015, 26.5% of households do not own a vehicle, where the average vehicles owned per household is 1.11 [20].

### Public Transit in Chicago
Chicago Transportation Authority (CTA) is the second largest transit agency in the U.S. as of 2018 [21]. The Chicago Transit Authority (CTA) runs and operates bus and rapid transit (rail) services within the city and the 35 surrounding suburbs. There are 1,864 buses that run 129 routes and 1,429 rail cars that serve 145 stations [22]. Additionally, CTA operates certain routes and lines during early morning and late-night hours, and some operate all hours of the day.

### Ride Hailing Services in Chicago
Historical data on the services present during our study period is unavailable at this time. However, as of January 2020, three personal-car ride-hailing services operate in the City of Chicago: Uber, Lyft, and Via [23].

### Data

#### TNP (Transportation Network Providers) – Trips Dataset
This dataset served as our primary source data for ride-hailing trips and was obtained from the City of Chicago's online data portal. The dataset contains 129 million unique TNC trips that span from November 2018 to the present day and is aggregated by the month [24]. We chose to only study one month, June 2019 because it does not contain any nationally recognized holidays that could hinder the representativeness of the results. Our selection from this dataset contains all TNC trips that have a trip start time on or after June 1, 2019 12:00:00 AM and before July 1, 2019 12:00:00 AM. The dataset contains 23 fields per trip, including a unique identifier, trip start and end time, pick-up and drop-off longitudinal and latitudinal coordinates, pick-up and drop-off census tract ID, trip fare, and if the ride was authorized as "shared" through the respective TNC app.

#### Public Transit Data (General Transit Feed Specification (GTFS) Dataset)
To perform a public transit network analysis in ArcGIS, it requires the GTFS dataset corresponding the area of interest. GTFS datasets are publicly available from OpenMobilityData, which contains archived real-time and fixed components of transit agencies' schedules [25]. The dataset holds the corresponding schedules, routes, stops, and transfers for the time period. With respect to this paper, this dataset will be integrated into ArcGIS such that the Network Analysis program can identify the corresponding transit route under spatial and temporal conditions.

### Preliminary Data Analyses
In Figure 3 below, is a preliminary analysis of the dataset in terms of trip passenger count and day type, weekday (M-F) or weekend (Saturday and Sunday). Over the month of June 2019, there were 8,136,461 ride-hailing trips serviced in the City of Chicago. This figure demonstrates that single occupancy ride-hailing trips have significantly greater demand than pooled trips. When compared to driving alone, this modality has a higher contribution towards congestion because its utility is comparably low due to deadheading mileage.

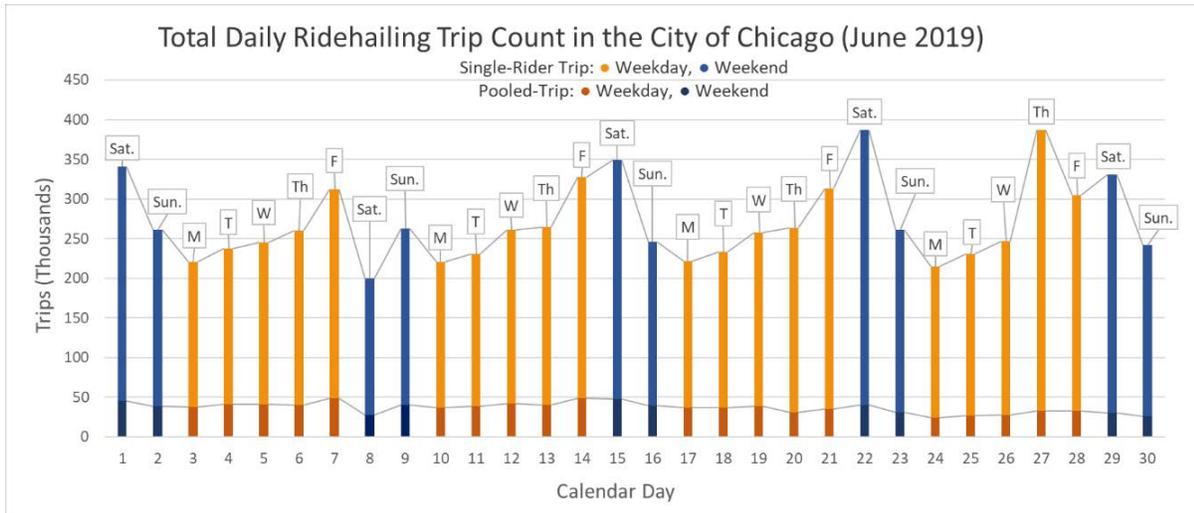

**Figure 3: Daily Ride-hailing Trip Counts in the City of Chicago during June 2019**

The next two figures show the distribution of trips by starting hour, by type of day; weekday trips (Monday-Friday) are depicted in Figure 4 and weekend trips (Saturday – Sunday) are depicted in Figure 5.

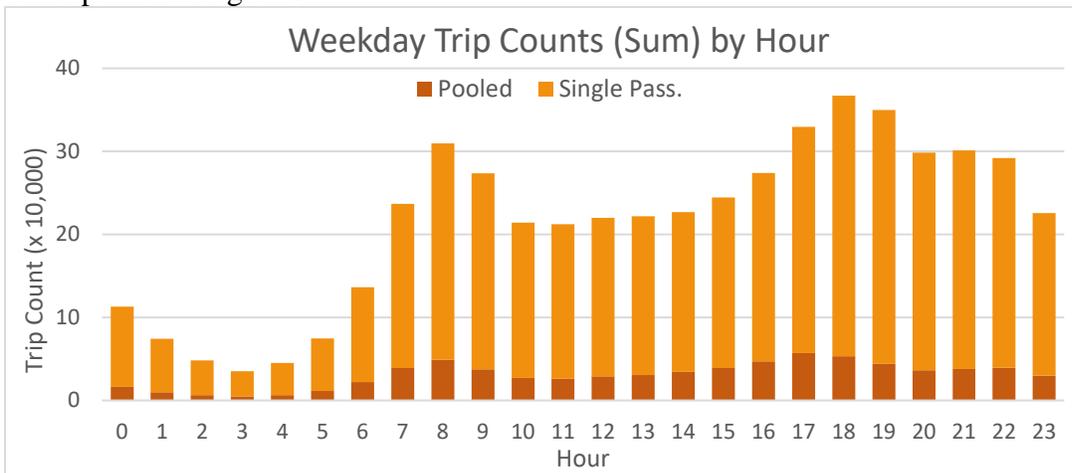

**Figure 4: Weekday Trip Counts (Sum) by Hour**

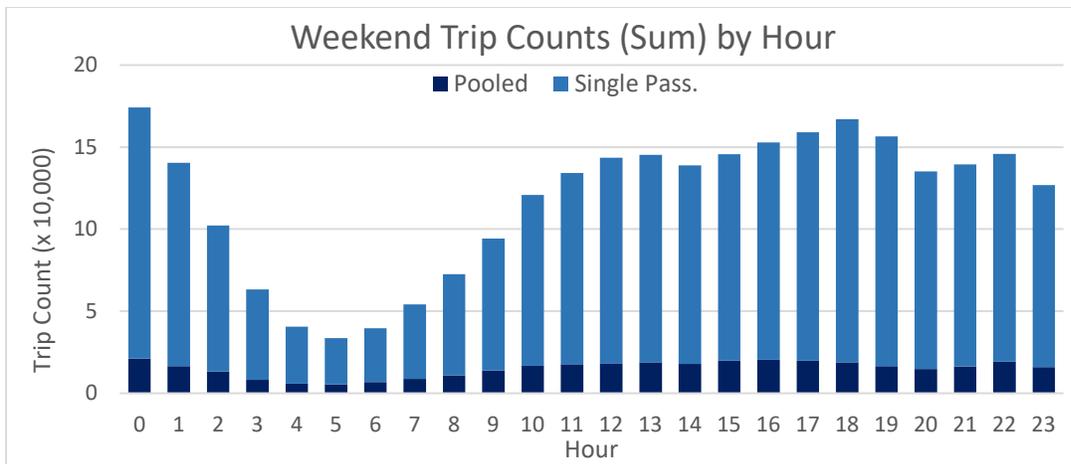

**Figure 5: Weekend Trip Counts (Sum) by Hour**

For weekday trips, two demand peaks exist at hours 8 and 16, whereas for weekend trips peak periods are not as distinct. On weekends, people tend to participate in social and leisure activities that are not confined to the traditional 9-5 work schedule. Hence, travel demand is more evenly distributed throughout daylight hours.

## Data Processing

Data processing was completed in three steps, with the ultimate output being the probability of a rider choosing public transit. This probability is derived from a multinomial nested logit (MNL) model based on the Chicago's travel behaviors in 2015 [11]. This model is descriptively explained, and its relevancy is introduced later in this section.

As a brief overview, the first two step were performed in the program, *ArcGIS*, using two separate tools: (1) Route Analyst and (2) Spatial Join. These two steps are novel, in that we combine GTFS data and source data to compute the time-conscious transit-equivalent route. The output of these two steps, per RH trip, were a transit-equivalent trip and the number of transfers required to complete the trip. For the third and final step, we used an existing utility model to compute the P(Transit|CTA), the probability of choosing to service a trip using CTA.

Considering the scope of our research, we chose to conduct a thorough literature review to find an existing utility model with obtainable input values, and that was derived from a sample with similar demographics and travel behaviors. We compared our methodology and dataset against these models to determine which existing utility model was most suitable. We selected the multinomial nested logit (MNL) model developed by Javanmardi et al. [11]. Their model was developed from a revealed preference survey using Google Maps API and RTA's Goroo TripPlanner trip data. The purpose of their MNL model is to measure the varying mode choice behaviors regarding alternative transportation, with increased accuracy from RP surveys. This model was developed from trips executed within the city limits of Chicago. Given the model shares the same study region of this paper, it is the most appropriate option in that it will capture the travel behavior with a greater degree of accuracy. Lastly, the study year (2015) of their research is appropriate in that ride-hailing was introduced to Chicago before that time, thus their model should capture any evolution of mode choice behavior and preferences towards or against alternative transportation.

The model's formulae are represented by the equations below (Equations 1-6). Given that the model was used for a range of modes, we modified its subscripts to align with the variables we used. The constant and coefficient values, and variables are mode-specific and were provided [11].

**Utility of Transit, $U_{Transit}$**

$$U_{Transit} = 2.93 - 1.04TT - 0.13TC - 0.17n_t - 0.77HHI + 0.45wrktrp - 0.39d_A - 023d_E \tag{1}$$

**Probability of Selecting Transit, $P_{Transit}$**

$$P_{Transit} = \frac{e^{U_{transit}}}{1 + e^{U_{transit}}} \tag{2}$$

**Utility of CTA, $U_{CTA}$**

$$U_{CTA} = -0.39TT - 0.059TC - 0.33n_t + 0.022n_{Stop,O} + 0.0089n_{Stop,D} + 0.77shptrp + 1.78wrktrp - 0.46HHI \tag{3}$$

**Probability of Selecting CTA, $P_{CTA}$**

$$P_{CTA} = \frac{e^{U_{CTA}}}{1 + e^{U_{CTA}}} \tag{4}$$

**Probability of Selecting CTA given Transit Selection, $P(Transit/CTA)$**

$$P_{CTA|Transit} = P_{Transit} \times P_{CTA} \tag{5}$$

The equations were executed in the respective order per trip, with Equation 6 outputting the final probability used in the analyses.

While this derivation of this model exhibited strong similarities to our study characteristics, it did contain several caveats. Our dataset did not completely satisfy all required input of the model, therefore we either generalized, estimated, or calculated these parameters. In doing so, assumptions were made. Table 1 outlines the input attributes, their corresponding definition, and their availability with respect to our dataset. Following the table, each assumption-based variable, and its calculation process(es) is explained in greater detail.

**Table 1: Utility Model Input Variables**

| Source | Variable | Definition |
|---|---|---|
| Output from spatial analysis | $TT$ | Total travel time (hr); wait time + walk time + transit time |
| Calculation; assumption-based | $TC$ | Total travel cost (USD); total cost of fare for transit trip |
| Spatial Analysis | $n_t$ | Number of transfers (transfers); |
| US Census Bureau | $HHI$ | Household income ($10^{-5}$ USD); |
| Calculation; assumption-based | $wrktrp$ | Purpose, Work trip (1/0); if trip purpose is for work, wrktrp = 1. Assumed if trip start day = weekday, and start hour in 5 a.m. – 7 p.m., trip purpose was for work |
| Calculation; assumption-based | $d_A$ | Access distance (km); walking distance from origin to first transit stop (pickup) |
| Calculation; assumption-based | $d_E$ | Egress distance (km); walking distance from last transit stop (drop-off) to destination |
| Spatial Analysis | $n_{Stop,O}$ | Number of transit stops in origin zone (stops); total number of transit stops within census tract containing origin |
| Spatial Analysis | $n_{Stop,D}$ | Number of transit stops in destination zone (stops); total number of transit stops within census tract containing destination |
| Calculation; assumption-based | $dest_{CBD}$ | Purpose, Destination in CBD during rush hour (1/0); if trip destination is within the geographic boundary of the Chicago CBD, and started during rush hour, $dest_{CBD} = 1$. |

**Household Income** *(HHI):* given the privatization of the ride-hailing dataset, we were not granted access to socioeconomic characteristics of the individual ride-hailer. However, we computed the HHI for a rider using a dataset containing the average HHI per census tract. We defined the HHI to equal the average HHI of the origin, if the trip was executed on a weekday between 5:00 AM and 12:00 PM, or of the destination, if the trip was executed on a weekday between 12:00 PM and 7:00 PM. For all trips outside this boundary, the HHI was defined as the average between the origin and destination HHI.

**Census Tracts of Origins and Destinations**: The ride-hailing dataset contained two fields for the origin and destination tract values, although a group of trips would had null origin or destination tract values but would have the geographic coordinates of the origin and destination. To account for this, we estimated the corresponding tract IDs via a minimum distance program in Matlab. First, we obtained the geographic boundaries of the census tracts from the Chicago Data Portal. We imported this SHP file into ArcGIS and performed geometry calculations to calculate the tract's centroid and to output the corresponding latitudes and longitudes. This output table was imported into Matlab as a matrix. Using each trip origin and destination latitude and longitude, we calculated the distance between each tract and O-D coordinates. The census tract ID corresponding to the smallest distance value was selected and replaced the null value for the origin or destination tract value.

**Number of Transit Stops per O-D Zone** *nStopOrigin; nStopDest*: these two attributes were a function of the O-D census tracts for the trip and were calculated as the number of transit stops in the corresponding origin or destination zone (census tract). For each trip, we indexed the number of stops in the O-D tracts from this table and appended these values to the final table for the utility calculation.

**Saturday/Sunday Classification**: To test true for the attributes below (*destCBD, wrktrp*), a trip could not have been serviced on a Saturday or Sunday. Therefore, to classify a 'weekend' (Saturday or Sunday) trip, we composed a vector of June 2019 calendar days corresponding to each pair of Saturdays and Sundays. If a trip's start calendar day was identified as a weekend day, then it results in a *false* value.

**Destination within CBD at Rush Hour (*destCBD*)**: Due to the anonymity of the dataset, we did not have individual details on trip characteristics such as the trip purpose. We assumed the AM and PM peak period to occur between 6:00:00 and 9:00:00 AM and 16:00:00 and 19:00:00 PM, respectively. For a trip to test "true" (destCBD = 1), we first determined if the destination census tract was a member of the CBD census tracts array. If true, the trip start hour was then tested against the two peak periods, otherwise, *destCBD* = 0.

**Trip Purpose: Work (*wrktrp*)**: like *destCBD*, we assumed that a trip was deemed as a "to-or-from work" commute (wrktrp = 1) if the start time lied between 5:00:00 and 19:00:00 on a weekday.

## Analysis

### Replaceability of a Transit Trip

Ultimately, to determine if a ride-hailing trip "replaced" its transit-equivalent trip, we compute the probability of using CTA. The magnitude of these probabilities indicates the viability of public transit serving a specified trip and depends on how favorable the trip's LOS attributes and trip-specific characteristics are to the rider. We chose to classify a transit-equivalent trip by its replaceability, categorized by two groups: replaced (R) trips and not-replaced (NR) trips. We initially assumed the threshold value distinguishing a trip being "replaced" (R) or "not replaced" (NR), to be 0.5. Where all trips with a P(Transit|CTA) < 0.5 were classified as not replaced and

all trips with a P(Transit|CTA)>=0.5 were classified as replaced. However, following the first sensitivity analysis trial, we determined that trips with P(Transit|CTA) close to 0.5 switch between the R- and NR-groups. These trips are fuzzy and are not reliable indicators of true mode-choice modeling behavior. Thus, we chose to implement a buffer, where trips with P(Transit|CTA) = (0.45-0.55) are removed and excluded from the summary statistics. This modification is represented by the conditional statement below.

For an individual ride-hailing trip, $T$,

$$replaceability\ group_T = \begin{cases} Not\ Replaced\ (NR), & 0 \leq P(Transit|CTA) \leq 0.45 \\ Replaced\ (R), & 0.55 \leq P(Transit|CTA) \leq 1.0 \\ Buffer\ Zone, & 0.45 < P(Transit|CTA) < 0.55 \end{cases}$$

### Sensitivity of P(Transit|CTA)

We then chose to conduct a parametric sensitivity analysis of P(Transit|CTA) with respect to the following decision variables:

1. Transit stops per census tract (StopsinTract)
2. Household income (HHI)
3. Total travel time (TTT)
4. Walk time (WT)

The $P(Transit|CTA)$ was recalculated under 20 sensitivity conditions, for each decision variable. Per variable, its observed value was adjusted in increments of 5%, ranging from -50% to +50%. Given that each variable was tested independently, there were a total of 80 trials. For each sensitivity condition, the algorithm was executed with the adjusted parameter and a new P(Transit|CTA) was output for all trips and averaged per group (R/NR).

Assuming that both groups share the same standard deviation, we can estimate $\sigma$ by calculating the pooled standard deviation, $s_p$, with the equation below. The pooled standard deviation for the observed and sensitivity condition data sets, for group R or NR, is:

$$s_p(group, sensitivity\ condition)$$
$$= \sqrt{\frac{[(n_{observed} - 1) * s_{observed}^2] + [(n_i - 1) * s_i^2]}{(n_{observed} + n_i) - 2}} \quad (6)$$

Where,
$n_{observed}$ = the number of trips in the observed group
$s_{observed}$ = standard deviation of the observed group
$n_i$ = number of trips in the sensitivity group
$s_i$ = standard deviation of the sensitivity group

It should be noted that $n_{observed}$ and $s_{observed}$ are fixed values under all sensitivity conditions.

To measure the level of influence and statistical relationship of each decision variable and the $P(Transit|CTA)$, we performed a two-tailed pooled t-test. Considering there is no overlap between the observed and sensitivity condition data, the two-tailed test was most suitable. We performed the t-test per variable, sensitivity condition, and trip group (R and NR). The relationship between the t-statistic and the critical value indicate whether we accept or reject the null hypotheses stated below:

$H_{Null\ (R)}$ = The $\bar{P}(Transit|CTA)$ of the observed $R$ group is not statistically different from the $\bar{P}(Transit|CTA)$ of the sensitivity $R$ group.

$H_{Null\,(NR)} = $ The $\bar{P}(Transit|CTA)$ of the observed $NR$ group is not statistically different from the $\bar{P}(Transit|CTA)$ of the sensitivity $NR$ group.

If the t-statistic is greater than the critical value (1.365), then we reject the null hypothesis and refer to the alternative hypothesis. The alternative hypothesis opposes the null by concluding that there is a statistically significant difference between the observed and the sensitivity condition data. Meaning, the influence of the decision variable on the $P(Transit|CTA)$ is expected to have an effect on the whole population, similar to the effect of the sensitivity condition.

The following equation was used to compute the t-statistic per variable and group for each sensitivity condition (Equation 8):

$$t_i = \left| \frac{\bar{P}_i - \bar{P}_{observed}}{\sqrt{\left(\frac{{s_i}^2}{n_i}\right) + \left(\frac{{s_{observed}}^2}{n_{observed}}\right)}} \right| \tag{7}$$

Where,

$i$ = sample trips of group $g$ (R or NR), decision variable $var$, and percent-change condition $\%\Delta$.

$\bar{P}_i$ = mean $P(Transit|CTA)$

$P_{observed}$ = mean $P(Transit|CTA)$ for group $g$ (R/NR), decision variable $var$, and percentage-change condition $\%\Delta$.

$s_i$ = mean $P(Transit|CTA)$ for group $g$ (R/NR), decision variable $var$, and percentage-change condition $\%\Delta$.

$s_{observed}$ = mean $P(Transit|CTA)$ for group $g$ (R/NR), decision variable $var$, and percentage-change condition $\%\Delta$.

$n_i$ = sample size for group $g$ (R/NR), decision variable $var$, and percentage-change condition $\%\Delta$.

$n_{observed}$ = mean $P(Transit|CTA)$ for group $g$ (R/NR), decision variable $var$, and percentage-change condition $\%\Delta$.

## RESULTS AND DISCUSSION

The following section introduces the results from the replaceability and sensitivity analyses for the four sensitivity variables: TTT, WT, HHI, and SiT.

### Replaceability Analysis (Probability of Selecting CTA)

The P(Transit|CTA) estimation is a function of the abovementioned procedures and their respective outputs. To reduce the paper's length, we will summarize each group in terms of size and P(Transit|CTA).. For ease of recall, the group for a trip, $T$, is categorized by the following conditional:

$$replaceability\ group_T = \begin{cases} Not\ Replaced\ (NR), & 0 \le P(Transit|CTA) \le 0.45 \\ Replaced\ (R), & 0.55 \le P(Transit|CTA) \le 1.0 \\ Buffer\ Zone, & 0.45 < P(Transit|CTA) < 0.55 \end{cases}$$

Of the 7,949,902 trips output from the route analysis, approximately 8% (646,808 trips) had a probability lying within the buffer range of 0.45<P<0.55. Moving forward, we will be summarizing findings in terms of the R and NR groups only.

## Sensitivity Analysis

Results from our sensitivity analyses were extensive and will be summarized for the submission guidelines of this paper.

For the aforementioned sensitivity parameters, t-tests were conducted at the 95% confidence level. Under each sensitivity condition, all variables exhibited t-values greater than the critical value. This implies that with 5% error, we can assume the adjustment of each of those four parameters will have a statistically meaningful impact on the P(Transit|CTA).

For each trip, the variable of interest was adjusted, the P(Transit|CTA) was recalculated, and the trip was recategorized based on its magnitude. Thus, it is important to consider that for each percent-change condition, the sample for R and NR trip groups will vary in size. Moreover, means of both groups will fluctuate with the sensitivity condition.

We will introduce stacked bar charts per sensitivity variable that depicts the overall weighted mean P and trip count per sensitivity-condition. Equation 10 below is used to compute the total weighted mean P(Transit|CTA):

$$P_{weighted} = P_R \left(\frac{n_R}{n_T}\right) + P_{NR} \left(\frac{n_{NR}}{n_T}\right) \tag{8}$$

The data labels (percentages) within each bar corresponds to the percentage of total trips ($n_T$) that each group contains. Per the legend, the blue portion of the stacked bar corresponds to the replaced (R) trips, whereas the orange portion corresponds to the not replaced (NR) trips. It should be clarified that these percentages are independent from the portion heights of the stacked bars.

In summary, the first three variables (*TTT*, *WT*, and *HHI*) exhibit an inverse correlation with P(Transit|CTA) whereas the remaining variable, *SiT*, has a positive relationship with P(Transit|CTA).

The first two variables are associated to travel time: *WT* and *TTT*. The P(Transit|CTA) is more sensitive to adjustments in *TTT* when compared to WT. The range of the weighted probability for *TTT* is approximately 0.20, as opposed to 0.10 for WT. The relatively heightened sensitivity for *TTT* can be explained by *TTT*'s formula and the nesting of *WT* in its value. Recall that the *TTT* is the sum of the *IVTT*, wait time, and walk time. Meaning, a 50% decrease in *TTT* includes a 50% decrease in *IVTT* and wait time in addition to a decrease in walk time. Whereas a 50% decrease in *WT* does not include the reduction in *IVTT* and walk time. When the new *TTT* is calculated, it uses the observed wait time and *IVTT*, but changes the *WT*. While these differences exist, they are relatively small in comparison to their distribution patterns and weighted probability values. In Figure 6 and Figure 7, the volumetric share between *R* and *NR* trips is indistinguishable.

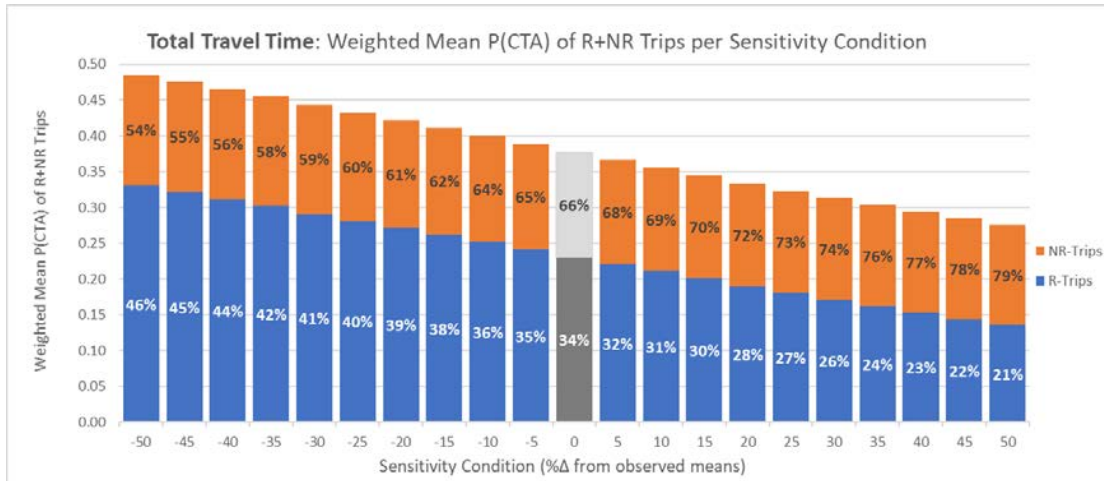

**Figure 6: Weighted Mean Probability and Trip Distribution between Groups per Condition for Sensitivity Variable: Total Travel Time (TTT)**

When analyzing the *WT*, it is important to consider that a trip's replaceability is contingent on the persons' capabilities of physical exertion. For example, an elderly, disabled person may need to traverse two blocks to get to the grocery store. Under ArcGIS' Route Analysis program it will likely output that walking is most efficient, although given the user's conditions, walking is not an option. Additionally, the replaceability is influenced by the user's safety, which is dependent upon the perception of the route's surrounding physical environment(s). These conditions produce a bias to act more conservatively such that hazardous events are mitigated. All of these conditions, concerns, and exceptions cannot be explicitly accounted for in our model.

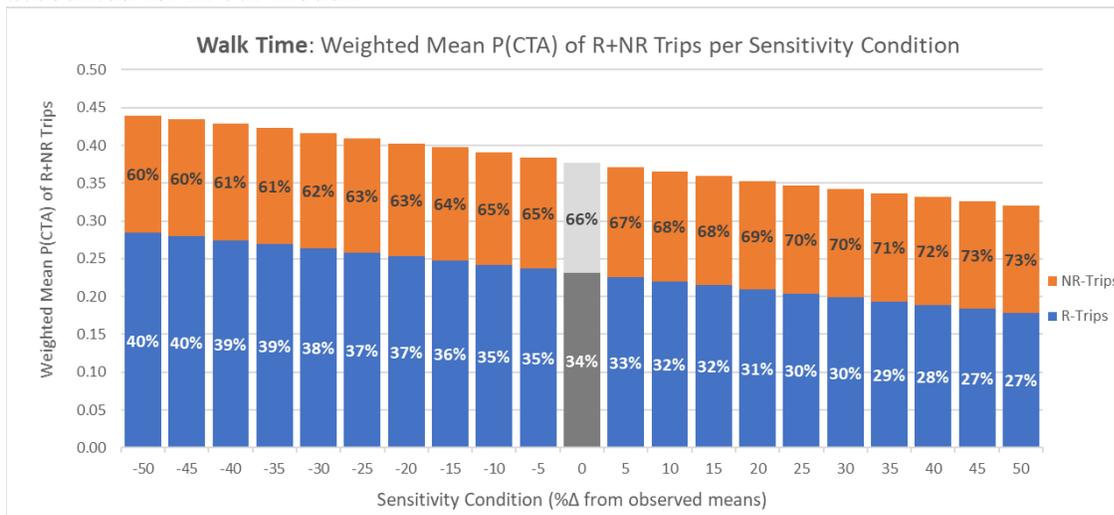

**Figure 7: Weighted Mean Probability and Trip Distribution between Groups per Condition for Sensitivity Variable: Walk Time (WT)**

The next variable, average *HHI*, has less impact on P(Transit|CTA) per Figure 8. In existing literature, it was determined that ride-hailers exhibit demographic characteristics that are at variance with the average American. Ride-hailers were found to be more educated and be of a

higher income class. Therefore, the use of the average *HHI* may undervalue that of the average ride-hailer and the accuracy of these results could be challenged. Nonetheless, the trend and behavior HHI has on P(Transit|CTA) is transposable.

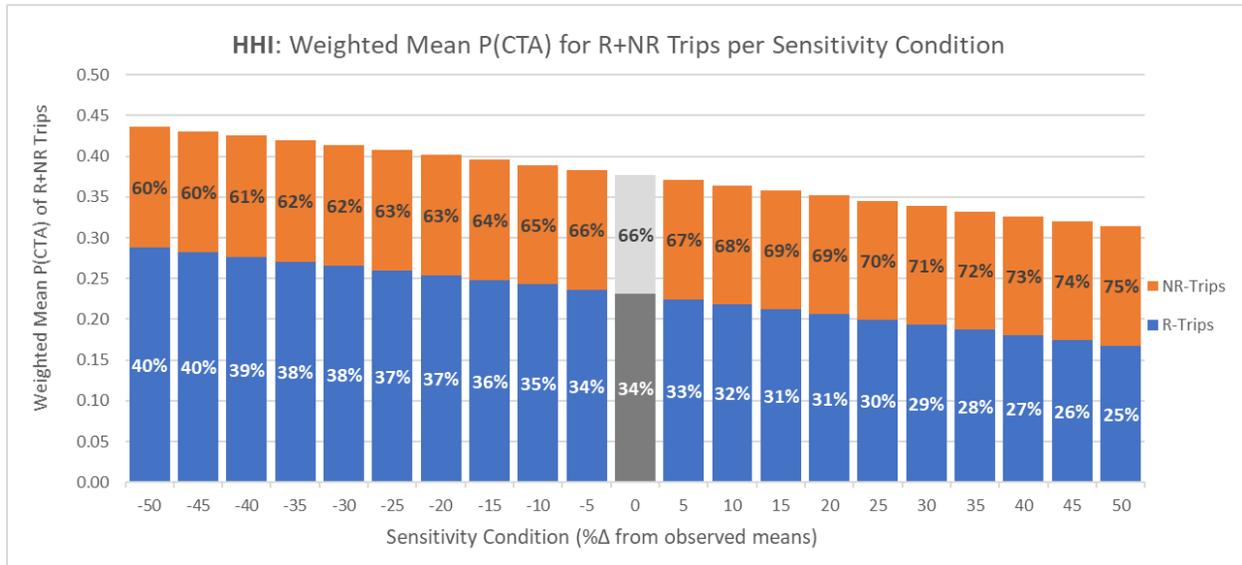

**Figure 8: Weighted Mean Probability and Trip Distribution between Groups per Condition for Sensitivity Variable: Average Household Income (HHI)**

In opposition to the prior variables, the number of transit stops per census tract (*SiT*) was exhibited a positive correlation with the P(Transit|CTA), as illustrated in Figure 9. The number of transit stops in a network has many implications on operations and ridership. An increase in transit stops implies an increase in route LOS. As the distance between consecutive stops is decreased, the average access and egress distance decreases. With this increase in accessibility, the volume of serviceable patrons increases.

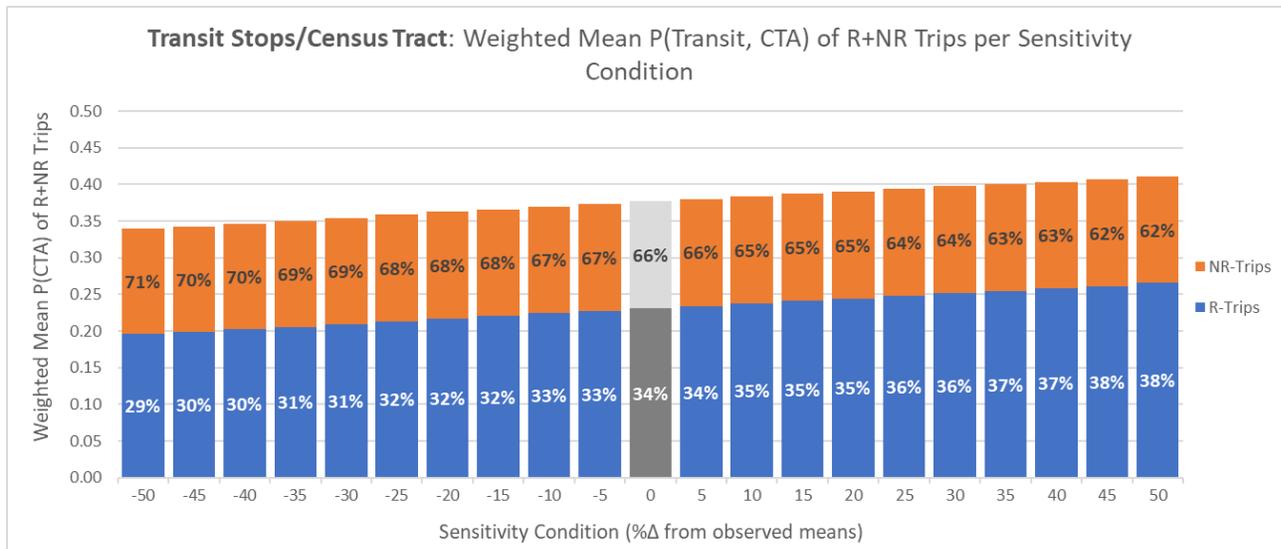

**Figure 9: Weighted Mean Probability and Trip Distribution between Groups per Condition for Sensitivity Variable: Transit Stops per Census Tract**

Although there exist caveats with the more extreme positive sensitivity conditions that are not captured in the utility model used. The addition of transit stops to existing routes must be optimized to account for added lost time. Delay has the opportunity to incur at every transit stop during the approach, boarding and dwelling, and exit. Boarding and dwell times can quickly accumulate during peak period hours when there are large platoons of entering riders, and there is discontinuity in payment forms. Additionally, this is a consequence of increased ridership. The second source is called the 're-entry' delay; this is the time required for the driver to merge into oncoming traffic. For every additional transit stop, one re-entry delay is incurred per cycle. The summation of these delays per stop and per cycle can adversely affect the travel time between stops, and the TTT of each rider. In summary, the addition of transit stops increases the accessibility and consequently, utility. Although designing addition to increase ridership must strategically consider implications it has on the existing travel times and LOS attributes.

## CONCLUSIONS

The impact of ride-hailing services on the recent decline in public transit ridership has not been widely explored. The current body of research is constrained to empirical studies that vary in methodologies used and analyze relatively small samples. To our knowledge, there are no studies that explore the research question using a massive dataset that contains individual trips over an extended period of time. Furthermore, our approach to exploring the research question is resourceful and novel. We define the replaceability of a RH trip by a series of spatiotemporal and mathematical analyses. First, the real-time transit equivalent trip is computed using the GTFS-integrated ArcGIS Route Analysis. Then, the probability of choosing transit over all other alternatives identified any potential viable transit-equivalent trips.

Our findings indicate that 31% of ride-hailing trips were executed where the transit alternative exhibited a competitive utility, with respect to travel times, fare/expenses, and workload. Consequently, over the month of June, the total revenue lost from trips replaced by ride-hailing is estimated to be $6,114,450[4]. If we assume the percentage of replaced trips and trip counts for each month can be represented by June 2019, then the total loss in fare revenue in Chicago over one year would be approximately 73 million dollars. Further, the ramifications of the demand transfer to ride-hailing services is not fully represented by the loss in revenue. As such, public transit agencies should employ strategies to increase transit utility such that a significant portion of this estimate can be recovered.

Publicly available ride-hailing trip data will likely maintain its anonymity by recording origins and destinations as their census tract centroids. Given it is unlikely for the precision to increase, studies that are macroscopic and encompass all attribute types (temporal, spatial, monetary) should be executed. However, the use of our methodologies and approach is only doable for regions that mandate the submission of all ride-hailing trips and their corresponding spatial and temporal parameters of the origin and destination. Recording and releasing this data will enable institutions to publish research that will provide a greater understanding of how ride-hailing impacts the transportation network and economy in varying geographic settings.

Moving forward, future research should focus on mode-choice behavior to thoroughly understand the factors favoring and disfavoring the use of transit. Replaced trips have comparable transit alternatives, hence there exists preferences and attitudes exclusive from LOS attributes that favor ride-hailing. Regarding NR trips, transit agencies should turn inwards and evaluate services, or the lack thereof, in the corresponding origin and destination zones.

---

[4] Estimated by multiplying the average fare ($2.26) of the replaced trips by the number of replaceable trips.

## AUTHER CONTRIBUTIONS

The work described in this article is the collaborative development of all authors, conceptualization, J.D.and H.A.R.; methodology, L.B., J.D., and H.A.R.; software, J.D. and L.B.; validation, L.B., J.D., and H.A.R.; formal analysis, L.B., J.D., and H.A.R.; investigation, L.B., J.D., and H.A.R.; writing—review and editing, L.B., J.D., and H.A.R.

## ACKNOWLEDGMENTS

This effort was funded by the Urban Mobility and Equity Center (UMEC).